\documentclass[aps,prd, twoside,superscriptaddress,showpacs, preprintnumbers, showpacs, nofootinbib,twocolumn]{revtex4-1}
\usepackage{amssymb,amsmath,latexsym,graphics, graphicx,epsfig,multirow,hyperref, verbatim} 

\usepackage{slashed}
\usepackage{subfigure}
\usepackage{feynmp}
\usepackage{graphicx}
\usepackage{amsfonts}
\usepackage{color}
\usepackage{cancel}
\usepackage[utf8]{inputenc}
\def\eq#1{{Eq.~(\ref{#1})}}
\def\eqs#1#2{{Eqs.~(\ref{#1})--(\ref{#2})}}
\def\fig#1{{Figure~\ref{#1}}}
\def\figs#1#2{{Figure~\ref{#1}--\ref{#2}}}

\def\tab#1{{Table~\ref{#1}}}

\def\hbar{\hspace{0pt}\raisebox{1pt}{$-$} \hspace{-7pt} h}

\def\5{\overline 5}

\definecolor{JJ}{RGB}{0,144,255}

\newcommand{\be}{\begin{equation}}
\newcommand{\ee}{\end{equation}}
\newcommand{\bea}{\begin{eqnarray}}
\newcommand{\eea}{\end{eqnarray}}

\newcommand{\ba}{\begin{eqnarray}}
\newcommand{\ea}{\end{eqnarray}}

\newcommand{\GeV}{\,\text{GeV}}
\newcommand{\TeV}{\,\text{TeV}}

\def\CtG{C_{tG}}

\begin{document}
%%%%%%%%%%%%%%%%%%%%%%%%%%%%%%%%%%%%%%%%%%%%%%%%%%%%%%%%%%%  FRONT PAGE
\title{Probing top-quark chromomagnetic dipole moment at next-to-leading order in QCD}

\author{Diogo Buarque Franzosi}
\affiliation{CP3-Origins, University of Southern Denmark, Campusvej 55, DK-5230 Odense M, Denmark}

\author{Cen Zhang}
\affiliation{ Department of Physics, Brookhaven National Laboratory, Upton, New
York 11973, USA }

\begin{abstract}
We present predictions at NLO accuracy in QCD for top-quark pair
production induced by an anomalous chromomagnetic dipole moment of the top
quark.  Our results are obtained for total as well as fully differential cross
sections, including matching to parton shower simulations.  This process is
expected to provide the most stringent direct limits on top-quark chromomagnetic
dipole moment.  We find that NLO corrections increase the contribution from the
dipole moment by about 50\% at the LHC, and significantly reduce the
renormalization and factorization scale dependence.  Using the NLO prediction,
we update the current limit from the Tevatron and the LHC measurements. 
Apart from total cross section, we also study other observables relevant for LHC
phenomenology.
\end{abstract}

%\keywords{}
%\pacs{11.10.Hi, 11.15.Ex,  12.39.Fe, 12.60.Fr}

%%%%%%%%%%%%%%%%%%%%%%%%%%%%%%%%%%%%%%%%%%%%%%%%%%%%%%%%%%%%%%%%%%%

\maketitle

\section{Introduction}\label{sec:introduction}

The top quark is expected to play an important role in new physics searches, due to
its large mass and strong coupling to the electroweak sector.   Strategies to
search for new physics effects in the top sector can be broadly divided into two
categories. In the first category, we search for new resonant states, such as
$t\bar t$ resonance and top partners. In the second category, new states are
assumed to be too heavy to be directly produced, and their indirect effects
are searched for in top-quark couplings.  

Currently, no new states have been discovered, and exclusion limits have been
placed, up to around several TeV scale for many new particles in either
complete or simplified models \cite{Agashe:2014kda}.  On the other hand, in the
second case the interaction of the top quark is becoming an ideal probe to new
physics.  On the experimental side, the millions of top quarks already produced
at the LHC together with the tens of millions expected in the coming years will
move top physics to a precision era.  Many detailed and accurate information on
various top-quark properties have been collected, and more will come.  On the
theory side, accurate SM predictions are also available, in general at next-to-next-to-leading order (NNLO) in
QCD and next-to-leading order (NLO) in electroweak for inclusive observables and at NLO in QCD for more
exclusive ones.  All of these provide the opportunity to extract or to
constrain different
anomalous top-quark couplings.  In this context, theoretical predictions
including QCD radiative corrections to anomalous top-quark couplings will be
necessary for extracting precise and reliable limits, as leading order (LO)
predictions in hadron colliders are often not reliable and suffer from large
scale uncertainties.  This has motivated a significant activity dedicated to
providing the NLO QCD corrections to top-quark
processes involving anomalous couplings, or higher-dimension operators
\cite{Drobnak:2010by,Drobnak:2010wh,Zhang:2013xya,Drobnak:2010ej,Zhang:2014rja,
Liu:2005dp,Gao:2009rf,Zhang:2011gh,Li:2011ek,Wang:2012gp,Shao:2011wa,
Degrande:2014tta,Rontsch:2014cca,Rontsch:2015una}.  However, NLO predictions
involving anomalous top-quark interactions are still far from complete.

In this work we focus on the chromomagnetic dipole moment (CMDM) of the top quark
in $t\bar t$ production.  The cross section of $t\bar t$ production
is one of the most accurately measured observables in top physics, and the effect
of an anomalous CMDM has been investigated in many studies
\cite{Stange:1993td,Li:1995fj,Yang:1995hq,Yang:1996dma,Martinez:2001qs,
Atwood:1994vm,Martinez:2007qf,Hioki:2009hm,Hioki:2010zu,Hioki:2013hva,
Kamenik:2011dk,Bernreuther:2013aga,Chatrchyan:2013wua,CMS:2014bea}. To the 
best of our knowledge, the contribution of top CMDM is known only at LO.
The goal of this work is to promote it to NLO including fully differential
productions and matching to parton showers, and investigate its impact on
the current limits of top-quark CMDM, as well as other observables, with either
stable or decayed $t\bar t$ system.  We shall mention that, apart from CMDM,
the top quark can also have an anomalous chromoelectric dipole moment (CEDM), which in
this work we will not discuss.  The reasons will be explained in the next
section.

At first glance, one might expect deviations induced
by anomalous top CMDM to be small, and therefore radiative corrections
of these contributions to be a higher-order effect.  However, in $t\bar t$
production at the LHC, the K factor from NLO QCD corrections is about 1.5 in
the SM, and is numerically not a higher-order effect.  If a similar K factor
applies to the CMDM contribution, it will be important to know the NLO
correction to the CMDM, so that a more accurate and stringent limit can be
obtained.  Moreover, this statement is too naive, since the issue is both on the
accuracy, i.e.~the central value, and on the precision, i.e.~the uncertainties
of a prediction.  LO predictions for processes at Hadron colliders always
suffer from large uncertainties due to scale variation, and the NLO predictions
are expected to significantly reduce these uncertainties and thus provide a
more reliable estimation of the possible range of the anomalous CMDM.  Finally,
in many cases NLO corrections can have an impact on kinematic distributions.
Knowing the accurate differential cross section from the CMDM is therefore
important in measurements where the shapes of the distributions are used.
We will show such examples, where LO prediction for the distributions does
not provide a reliable description.

The paper is organized as follows.  In section \ref{sec:2} we briefly discuss
the theoretical background of top CMDM. In section \ref{sec:3} we describe the
framework of our calculation and how it is implemented.  Our results 
for total cross sections and limits are presented in section \ref{sec:4}.
In section \ref{sec:5} we show several examples of exclusive distributions.
Section \ref{sec:6} is our conclusion.

\section{Theoretical background}
\label{sec:2}

The top-quark chromomagnetic and chromoelectric dipole moments, CMDM and CEDM,
can be parameterized by adding an effective term to the top-gluon coupling:
\begin{equation}
  \mathcal{L}_{ttg}=g_s\bar t\gamma^\mu T^A tG_\mu^A
  +\frac{g_s}{m_t}\bar t\sigma^{\mu\nu}\left( d_V+id_A\gamma_5 \right)T^AtG_{\mu\nu}^A
  \label{eq:eq1}
\end{equation}
where $g_s$ is the strong coupling, and $G_{\mu\nu}^A$ is the gluon field strength
tensor.  $d_V$ and $d_A$ in the second term represent the CMDM and
CEDM of the top quark respectively.  

The CMDM of the top quark can arise from various models of new physics.  The
Yukawa corrections to $gt\bar t$ vertex in two Higgs doublet model (2HDM) was
first considered in Ref.~\cite{Stange:1993td}, while the supersymmetric QCD and
electroweak corrections have been studied in
Ref.~\cite{Li:1995fj,Yang:1995hq,Yang:1996dma}.  Explicit expressions for CMDM
in 2HDM and in minimal supersymmetric standard model were given in
Ref.~\cite{Martinez:2001qs}. The top CMDM also arises quite naturally in
composite models and technicolor models \cite{Atwood:1994vm}.  For a more
general discussion of top CMDM in new physics scenarios we refer to
Ref.~\cite{Martinez:2007qf}.  Finally, a top CMDM operator may be loop-induced
by operator mixing effects, from other higher-dimensional operators generated
at higher scales.  An example can be found in Ref.~\cite{Cho:1994yu}.

At the LHC, CMDM is mainly constrained by $t\bar t$ production.  Direct limits
have been derived by previous studies
\cite{Hioki:2009hm,Hioki:2010zu,Hioki:2013hva,Kamenik:2011dk,Bernreuther:2013aga,
Chatrchyan:2013wua,CMS:2014bea}.  However, the contribution of top CMDM has
been known only at LO accuracy.  Our aim is to provide the NLO prediction, as
well as to study its impact on the total cross section and various distributions.

To go beyond LO calculation, a theoretical framework based on the dimension-six
Lagrangian of the SM is required.  This framework contains a complete set of
operators satisfying the symmetries of the SM, i.e.~the Lorentz symmetry
and the $SU(3)_C\times SU(2)_L\times U(1)_Y$ gauge symmetries.
It provides an unambiguous prescription for operator renormalization,
and thus allows for a complete and consistent treatment of the higher-order
corrections to the operators.  The Lagrangian including dimension-six operators
can be written as
\begin{equation}
  \mathcal{L}_{\rm EFT}=\mathcal{L}_{SM}+\sum_i \frac{C_iO_i}{\Lambda^2}+h.c.
\end{equation}
where $\Lambda$ is the scale of new physics.  In this work we work up to order
$\mathcal{O}(\Lambda^{-2})$, as going beyond this order would require
complete knowledge of dimension-eight operators.

\begin{figure}[t!] 
\begin{center}
 \includegraphics[width=.99\columnwidth]{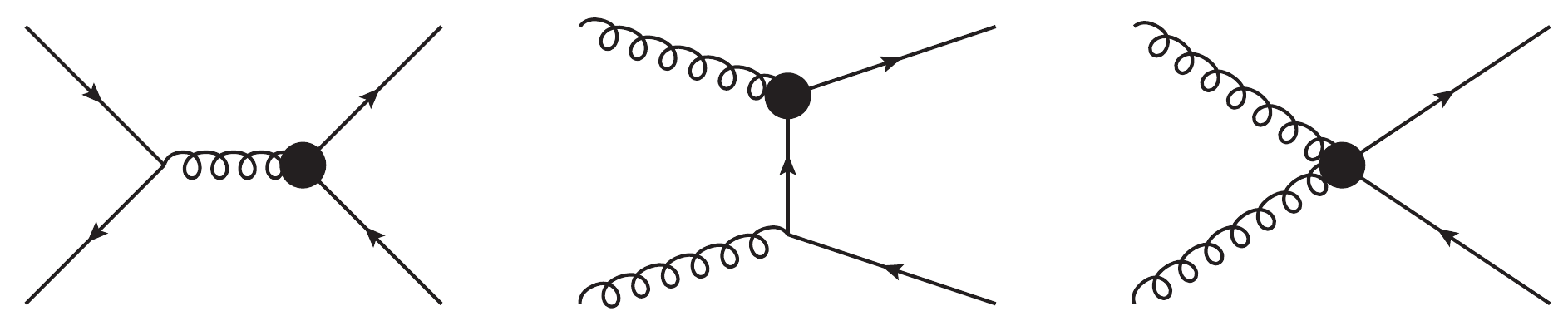}
 \caption{Representative tree-level diagrams of $t\bar t$ production with an effective
	 vertex form the operator $O_{tG}$.  Black dot represents effective
	 vertex from $O_{tG}$.}
\label{fig:diagLO}
\end{center}
\end{figure}

The top-quark CMDM in this framework is represented by a dimension-six operator
\begin{equation}
  O_{tG}=y_tg_s\left( \bar Q\sigma^{\mu\nu}T^At \right)\tilde\phi G_{\mu\nu}^A\ ,
\end{equation}
where $Q$ is the left-handed top- and bottom-quark doublet, $t$ the right-handed top,
$\phi$ the Higgs doublet, and $y_t$ the Yukawa coupling of the top quark.
$\tilde\phi=i\sigma^2\phi$.
This operator, after the electroweak symmetry breaking, takes the form of the
second term in \eq{eq:eq1}.
The relation between $d_V$ and the real part of the coefficient of $O_{tG}$
is given by
\begin{equation}
  d_V=\frac{ \mathrm{Re}C_{tG}m_t^2}{\Lambda^2}
\end{equation}
The operator $O_{tG}$ contributes to $t\bar t$ production at tree level by
modifying the standard $gt\bar t$ vertex, as well as inducing a new $ggt\bar t$
vertex, as shown in \fig{fig:diagLO}.  The effects of this operator in
top-quark processes at LO in QCD have been discussed in
Refs.~\cite{Zhang:2010dr,Degrande:2010kt}.

On the other hand, $d_A$, the CEDM, corresponds to the imaginary part of
$C_{tG}$. In this work, however, we are going to focus only on the CMDM.  This
is because the analysis of the CEDM at NLO follows a completely different
approach.  As we have mentioned above, in an approach based on the
dimension-six Lagrangian, we can only work up to order
$\mathcal{O}(\Lambda^{-2})$, and thus only the interference between the CEDM
and the SM amplitudes can be included.  At this order the contribution vanishes
in $t\bar t$ process because of the CP-odd nature of the CEDM, unless one
incorporates the decay of the top quarks.  As we shall see, our work is based
on the {\sc MadGraph5\_aMC@NLO} framework \cite{Alwall:2014hca}, where the 
spin correlation and the off-shellness of the top quark pairs
are simulated by using the {\sc MadSpin} package \cite{Artoisenet:2012st},
which is based on LO evaluation of the complete matrix element including top
decays.  Therefore it is not a suitable framework for the NLO corrections to
the CP-odd effects of the CEDM, and we will leave the NLO analysis of the
CEDM to future works.  Throughout the paper, we assume $C_{tG}$ to be real.  

When going to NLO in QCD, one needs to take into account the operator mixing
effects between $O_{tG}$ and other dimension-six operators that could give a
contribution to the same process at tree level.  For $t\bar t$ production,
these operators are
$O_{G}=g_sf^{ABC}G_{\mu}^{A\nu}G_{\nu}^{B\rho}G_{\rho}^{C\mu}$, $O_{\phi G}
=g_s^2\left( \phi^\dagger\phi \right)G^A_{\mu\nu}G^{A\mu\nu}$, and several
four-fermion operators \cite{Zhang:2010dr}.  It turns out that in $t\bar t$
production, the mixing from $O_{tG}$ to these operators is not relevant.
First of all, $O_{tG}$ does not mix into $O_{G}$ and four-fermion operators
\cite{Cho:1994yu}, because $O_{tG}$ is essentially a dimension-five operator 
if the Higgs field always takes the vacuum expectation value, which is always
true at the order we are working at. Second, $O_{tG}$ does mix into $O_{\phi G}$
\cite{Degrande:2012gr}, but such effects correspond to a $\mathcal{O}(y_t^2)$
correction to the LO process, and therefore of higher-order.  Finally,
the operators $O_{\phi G}$ and $O_{G}$ do mix into $O_{tG}$
\cite{Alonso:2013hga}, however it is consistent to assume that they vanish at all
scales, given that they are not renormalized by $O_{tG}$.  Therefore, as a first
step, in this work we assume $O_{tG}$ is the only non-vanishing operator,
and neglect other operators.  Note, however, that for a fully consistent
phenomenological study, the complete operator set must be included. 
A first example of the global approach was presented for the flavor-neutral
interactions of the top quark \cite{Durieux:2014xla}.  The NLO
predictions for other operators will be left to future works.

\section{Framework and implementation}
\label{sec:3}

\begin{figure}[t!] 
\begin{center}
 \includegraphics[width=.99\columnwidth]{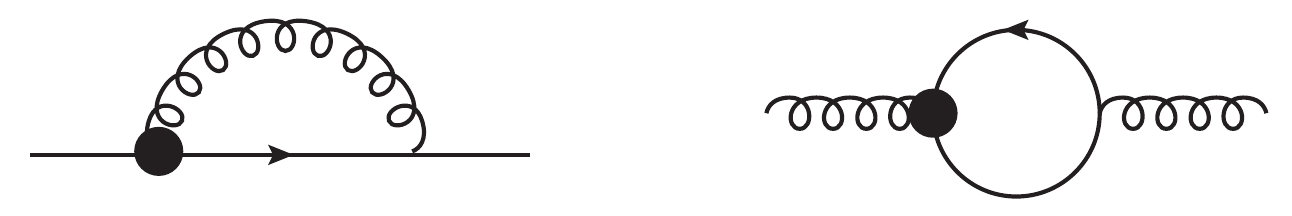}
 \caption{Contribution of $O_{tG}$ operator to top-quark and gluon wave functions.
 Black dot represents effective vertex from $O_{tG}$.}
\label{fig:diagWF}
\end{center}
\end{figure}

In an NLO calculation one has to choose a renormalization scheme.  Our scheme
is consistent with Ref.~\cite{Zhang:2014rja}.  For the SM part, we adopt
$\overline{MS}$ with five-flavor running in $\alpha_s$ with the top-quark
subtracted at zero momentum transfer \cite{Collins:1978wz}.  The bottom quark
mass is neglected.  Masses and wave-functions are renormalized on shell.
The dimension-six operator $O_{tG}$ then gives additional contributions to
top-quark and gluon fields renormalization, as shown in \fig{fig:diagWF}.  We
find
\begin{flalign}
  \delta Z_2^{(t)}&=\delta Z_{2,SM}^{(t)}-C_{tG}\frac{2\alpha_s m_t^2}{\pi\Lambda^2}
  D_\varepsilon \left( \frac{1}{\varepsilon_{UV}}+\frac{1}{3} \right)
  \label{eq:zt}
  \\
  \delta m_t &= \delta m_{t,SM} - C_{tG}\frac{4\alpha_s m_t^3}{\pi\Lambda^2}
  D_\varepsilon \left( \frac{1}{\varepsilon_{UV}}+\frac{1}{3} \right)
  \label{eq:zmt}
  \\
  \delta Z_2^{(g)}&=\delta Z_{2,SM}^{(g)}-C_{tG}\frac{2\alpha_s m_t^2}{\pi\Lambda^2}
  D_\varepsilon \frac{1}{\varepsilon_{UV}}\ ,
\end{flalign}
where
\begin{equation}
  D_\varepsilon\equiv \Gamma(1+\varepsilon)\left( \frac{4\pi\mu^2}{m_t^2}
  \right)^\varepsilon\ ,
\end{equation}
and $\mu$ is the renormalization scale.  In addition, the strong coupling counterterm,
$Z_{g_s}$, also gets a dimension-six contribution:
\begin{equation}
  \delta Z_{g_s}=\delta Z_{g_s,SM}+C_{tG}\frac{\alpha_s m_t^2}{\pi\Lambda^2}
  D_\varepsilon \frac{1}{\varepsilon_{UV}}\ ,
\end{equation}
that is to say the top-loop contribution with the operator $O_{tG}$ is also decoupled
from the running of $\alpha_s$, in the same way as in the SM.  Finally, for
operator coefficient we use $\overline{MS}$ subtraction.  The counterterm of
$C_{tG}$ is
\begin{equation}
  \delta Z_{C_{tG}}=\frac{\alpha_s}{6\pi}\Gamma(1+\varepsilon)(4\pi)^\varepsilon
  \label{eq:zc}
\end{equation}
This will lead to the running of $C_{tG}$.

One remark on \eq{eq:zmt} is in order.  Naively if $\overline{MS}$ is applied
to the complete set of dimension-six operators, one would expect that the
dimension-six ``Yukawa'' operator, $O_{t\phi}=y_t^3(\phi^\dagger\phi)(\bar
Qt)\tilde\phi$, will be renormalized by $O_{tG}$ and will be providing the UV
pole in the mass counterterm in \eq{eq:zmt}.  The remaining finite term,
however, still needs to be subtracted by introducing the mass counterterm.
Operator $O_{t\phi}$ does not have a physical effect in this process, as it
only shifts the top-quark mass which is an input parameter.  Therefore it is
equivalent to redefine $O_{t\phi}$ as $y_t^3(\phi^\dagger\phi-v^2/2)(\bar
Qt)\tilde\phi$, and shift the renormalization of the dimension-four component
of $O_{t\phi}$, i.e.~$-m_t^2/\Lambda^2y_t(\bar Qt)\tilde\phi$, to \eq{eq:zmt}.
This is more convenient since $O_{t\phi}$ then completely drops out from the
calculation.

Our calculation is performed using the {\sc MadGraph5\_aMC@NLO} framework
\cite{Alwall:2014hca}.  The operator $O_{tG}$ is implemented in the UFO format
\cite{Degrande:2011ua} by using the {\sc FeynRules} package
\cite{Alloul:2013bka}. Helicity amplitude routines are generated by {\sc ALOHA}
\cite{deAquino:2011ub}.
The evaluation of the loop corrections requires two additional 
pieces, the UV counterterms and the rational R2 terms which are required
by the OPP technique \cite{Ossola:2006us}.
The UV counterterms are computed according to \eqs{eq:zt}{eq:zc},
while the R2 terms are generated by the NLOCT package
\cite{Degrande:2014vpa}.  The calculation is then automatically performed by
{\sc MadGraph5\_aMC@NLO} at NLO accuracy, and matched to parton shower
via the MC@NLO formalism \cite{Frixione:2002ik}.

Several checks of the implementation have been done, including the 
gauge invariance of all virtual contributions, and the pole cancellation
when combining virtual and real contributions. In addition, we checked that
all relevant UV and R2 terms are correctly implemented, by computing individual
diagrams with {\sc MadLoop} \cite{Hirschi:2011pa} and comparing with analytical
results obtained by using {\sc FormCalc} and {\sc LoopTools}
\cite{Hahn:1998yk}.

\section{Total cross section}
\label{sec:4}

In this section we give the NLO total cross section from $O_{tG}$, and place
limits on its size, using available measurements from the Tevatron and the LHC.

As mentioned above, we work up to $\mathcal{O}(\Lambda^{-2})$,
which means we insert in each diagram at most one
effective vertex from $O_{tG}$.  The total cross section then becomes a quadratic
function of $C_{tG}/\Lambda^2$, 
\be \sigma =
\sigma_{\rm SM}+\frac{\CtG}{\Lambda^2}\beta_1+\left(\frac{\CtG}{\Lambda^2}\right)^2
\beta_2\,.  \label{eq:betaxs} \ee 
The $\beta_1$ term represents the contribution from $O_{tG}$ at order
$\mathcal{O}(\Lambda^{-2})$.
The quadratic $\beta_2$ term, on the other hand, does not have a physical
meaning without a complete calculation at $\mathcal{O}(\Lambda^{-4})$, and needs
to be dropped.  However the size of this term can be used to gauge the range
in which the approach itself is valid, or in other words, the expansion in
$1/\Lambda$ converges.

To extract $\beta_{1,2}$, we perform the calculation with $C_{tG}$ taking
different values: $0,\pm1,\pm2$, and fit the resulting cross sections to
\eq{eq:betaxs}.  Each run is performed with 9 combinations of
($\mu_R$,$\mu_F$), where $\mu_R$ is the renormalization scale and $\mu_F$ the
factorization scale, each can take values $\mu/2$, $\mu$ and $2\mu$, with the
central value $\mu=m_t$.  This allows us to extract the scale variation of
$\beta_1$.  In our calculation $m_t=173.3$ GeV, and we use the NNPDF 2.3 set
of the parton distribution functions \cite{Ball:2012cx}.  The values of
$\beta_{1,2}$, both at LO and NLO, are given in \tab{tab:beta} for Tevatron,
LHC 8 TeV, LHC 13 TeV and LHC 14 TeV runs. A significant improvement in the
scale dependence can be noticed. A large K factor is found at the LHC. The
sizes of $\beta_2$ implies that the effective approach is valid given that
$C_{tG}/\Lambda^2\lesssim1 $ TeV$^{-2}$. Our LO results agree with
Ref.~\cite{Aguilar-Saavedra:2014iga} once we take into account scale variation
and note the opposite sign convention of $d_V$. 

\begin{table}
\begin{tabular}{|l|c|c|c|}
\hline
\hfill$\beta_1$ & LO [pb TeV$^{2}$] & NLO [pb TeV$^{2}$] &  K factor\\ \hline
Tevatron   & $1.61^{+0.66}_{-0.43}\,\, ^{(+41\%)}_{(-27\%)} $	
&  $1.810^{+0.073}_{-0.197}\,\, ^{(+4.05\%)}_{(-10.88\%)} $	 &  1.12\\
\hline
LHC8	& $50.7^{+17.3}_{-12.4}\,\, ^{(+34\%)}_{(-25\%)} $	
&  $72.62^{+9.26}_{-10.53}\,\, ^{(+12.7\%)}_{(-14.5\%)} $	& 1.43\\
\hline
LHC13	& $161.6^{+48.0}_{-36.2}\,\, ^{(+29.7\%)}_{(-22.4\%)} $
&  $239.5^{+29.0}_{-31.8}\,\, ^{(+12.1\%)}_{(-13.3\%)} $	& 1.48\\
\hline
LHC14	& $191.3^{+55.6}_{-42.2}\,\, ^{(+29.0\%)}_{(-22.0\%)} $
&  $283.0^{+33.6}_{-36.9}\,\, ^{(+11.9\%)}_{(-13.1\%)} $	& 1.48\\
\hline
\end{tabular}
\\[5pt]
\begin{tabular}{|l|c|c|}
	\hline
	\hfill$\beta_2$ & LO [pb TeV$^4$] & NLO [pb TeV$^4$]
	\\\hline
	Tevatron & 0.156 & 0.158
	\\\hline
	LHC8 & 8.94 & 11.8
	\\\hline
	LHC13 & 30.0 & 43.2
	\\\hline
	LHC14 & 35.7 & 51.6
	\\\hline
\end{tabular}
\caption{Values of $\beta_1$ and $\beta_2$ at LO and NLO precisions for the
	Tevatron, and for the LHC at 8, 13, and 14 TeV.  The respective
	K factors for the central values of $\beta_1$ are also
	shown.\label{tab:beta}}
\end{table}

With these results we can set bounds on the size of $O_{tG}$ using total cross
section measurements.  We replace the $\sigma_{\rm SM}$ in \eq{eq:betaxs} by the most
precise SM predictions at NNLO+NNLL accuracy in QCD, which are $\sigma^{\rm
TeV}_{\rm SM}=7.148\pm 0.218 \text{ pb}$ and $ \sigma^{\rm LHC}_{\rm SM}=244.9
\pm9.7 \text{ pb}$ respectively for Tevatron and for LHC at 8 TeV
\cite{Czakon:2013goa}.  We sum the scale and PDF uncertainties in quadrature
and symmetrize the error around a central value. 
The combined measurement at the Tevatron (LHC) is
$\sigma^{\rm TeV}_{\rm exp}=7.51\pm 0.40 \text{ pb}$
($\sigma^{\rm LHC}_{\rm exp}=240.6 \pm8.5 \text{ pb}$)
\cite{Aaltonen:2013wca,CMS-PAS-TOP-14-016},
where we have corrected for the top-mass difference using the prescription given
in these references.
The value of $C_{tG}$ can be extracted using
\begin{equation}
  \frac{C_{tG}}{\Lambda^2}=\frac{\sigma_{\rm exp}-\sigma_{\rm SM}}{\beta_1}\,,
\end{equation}
together with its corresponding uncertainty, given by
\begin{equation}
  \frac{\delta
    C_{tG}}{\Lambda^2}=\left|\frac{\sigma_{\rm exp}-\sigma_{\rm SM}}{\beta_1}\right|
  \left[
    \frac{\epsilon_{\rm exp}^2+\epsilon_{\rm SM}^2}{(\sigma_{\rm exp}-\sigma_{\rm SM})^2}
    +\left( \frac{\epsilon_{\beta_1}}{\beta_1} \right)^2
    \right]^{\frac{1}{2}}
    \label{eq:err}
\end{equation}
where the experimental error $\epsilon_{\rm exp}$ and the theoretical NNLO error
$\epsilon_{\rm SM}$ are summed in quadrature, and $\epsilon_{\beta_1}$ is the 
(symmetrized) error of $\beta_1$ due to scale variations.  We assume no
correlation between the SM NNLO prediction and $\beta_1$.
One could also add in \eq{eq:err} a term representing the error from the missing
$\mathcal{O}(\Lambda^{-4})$ terms, which can be estimated using
$\beta_2(C_{tG}/\Lambda^2)^2$, but the changes in the limits are negligible.

We show the 95\% CL allowed region for $C_{tG}$ in \tab{tab:ctgbounds}. The
improvement of NLO calculation for Tevatron is mild due to the small K factor.
In the LHC cases the allowed range is significantly reduced at NLO. We also
give the expected limit at the LHC 14 TeV run, assuming an experimental
error of $\pm5\%$.

\begin{table}
\begin{tabular}{|l|c|c||c|c|c|}
\hline
& LO [TeV$^{-2}$]  & NLO [TeV$^{-2}$] \\
\hline
Tevatron& [-0.33, 0.75]	   &  [-0.32, 0.73]	\\
\hline
LHC8& [-0.56, 0.41]  &  [-0.42, 0.30]	\\
\hline
LHC14& [-0.56, 0.61]     &  [-0.39, 0.43]\\
\hline
\end{tabular}
\caption{Limits on $C_{tG}/\Lambda^2$. The corresponding limits 
combining Tevatrion and LHC8, in terms of
$d_V$, is $[-0.0099,0.0123]$ at LO and $[-0.0096,0.0090]$ at NLO (note the
opposite sign convention of $d_V$ in \cite{Aguilar-Saavedra:2014iga}).
For LHC14 we assume a $5\%$ experimental error.}
\label{tab:ctgbounds}
\end{table}

\section{Distributions}
\label{sec:5}

Our calculation is implemented via the {\sc MadGraph5\_aMC@NLO} framework,
therefore simulation of any observable is automatic. In this
section we present a few representative distributions of variables of
particular relevance for LHC phenomenology.  To simulate parton shower we have
used the {\sc Herwig 6} code \cite{Corcella:2000bw}.  Other shower programs are also
available, including {\sc Herwig++} \cite{Bellm:2013lba} and {\sc Pythia 8}
\cite{Sjostrand:2007gs}.

\subsection{Stable top quarks}

\begin{figure}[t!] 
\begin{center}
 \includegraphics[width=.99\columnwidth]{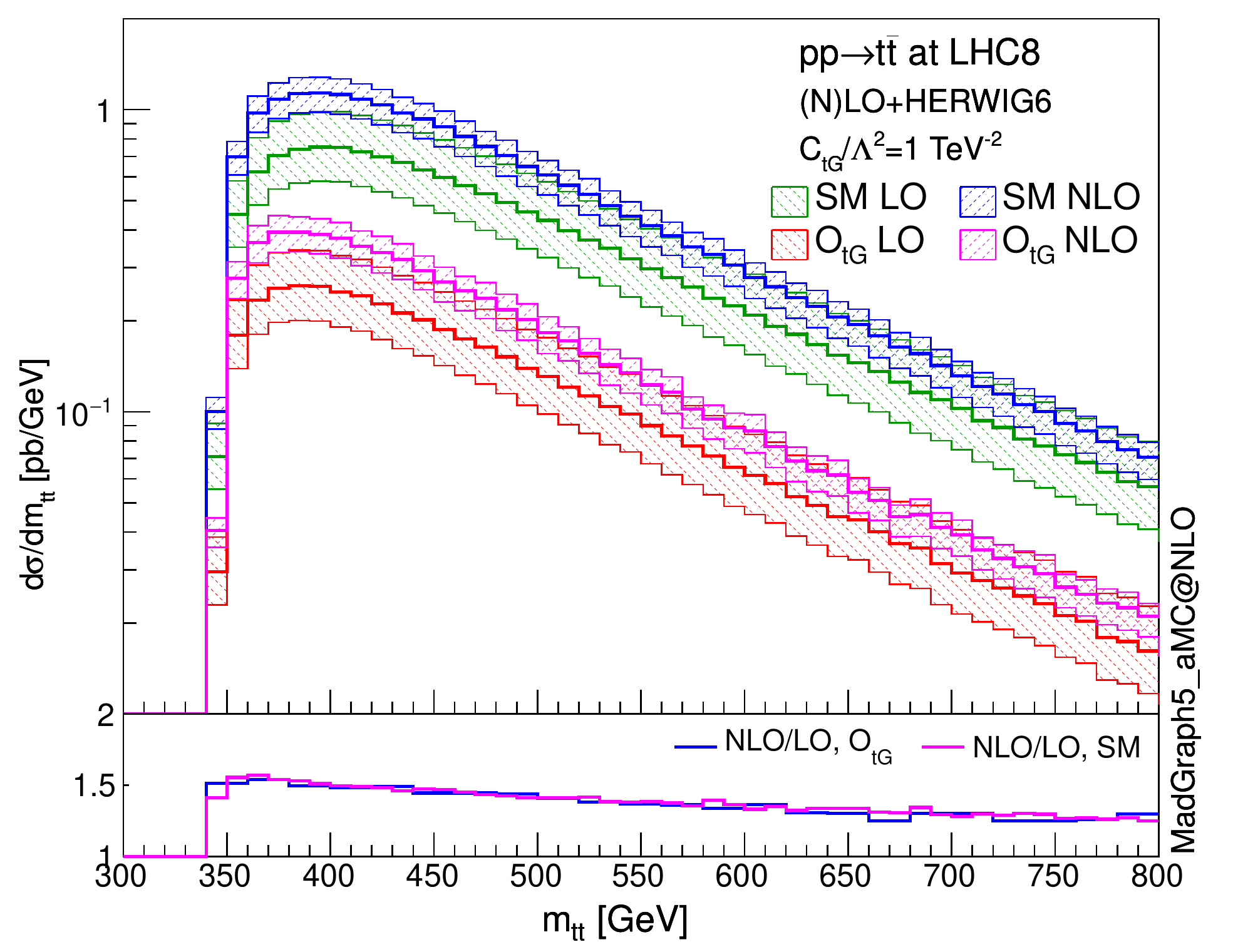}
 \caption{Top quark pair invariant mass distribution at LHC 8 TeV.}
\label{fig:mtt8}
\end{center}
\end{figure}

\begin{figure}[t!] 
\begin{center}
 \includegraphics[width=.99\columnwidth]{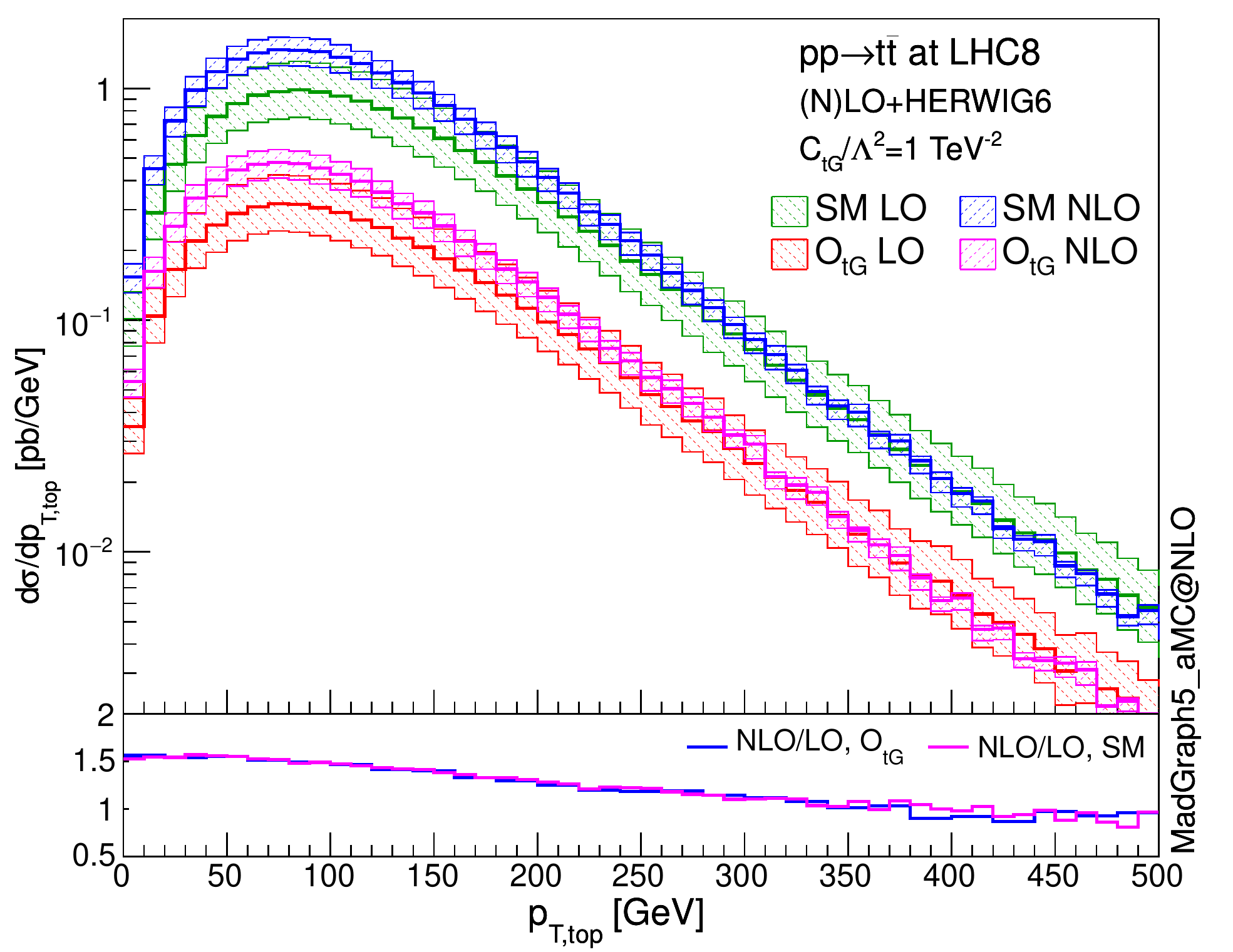}
 \caption{Top quark transverse momentum distribution at LHC 8 TeV.}
\label{fig:pt_t}
\end{center}
\end{figure}

We first look at kinematic observables constructed from stable top-quark pairs.
In \fig{fig:mtt8} and \fig{fig:pt_t} we show the invariant mass of the top
anti-top system and the transverse momentum of the top-quark for the LHC at
8 TeV, in each case with the differential K factor displayed in the lower
panel.  The contribution from $O_{tG}$ is extracted by generating event samples
with $C_{tG}=\pm2$ separately and taking the difference, in order to get rid of
the quadratic terms in $C_{tG}$.  These observables can serve as discriminators
in case any deviation form the SM is observed, and will be useful in
determining the type of new physics \cite{Degrande:2010kt}.  One can see that
the NLO computation reduces the scale variation.  The differential K factor is
not a constant and drops at higher scales, however in both distributions we
observe that the K factor of the $O_{tG}$ contribution is similar to that of the SM
contribution, so using the SM K factor to rescale the LO event samples from $O_{tG}$
can be a good approximation of the complete NLO result.  Note that we have
chosen $C_{tG}/\Lambda^2=1$ TeV$^{-2}$ for convenience, even though this value
is already excluded by the current limits.  One can always rescale the curve to
get corresponding result for any other value of $C_{tG}$, as we have already
removed the quadratic dependence on $C_{tG}$.

\begin{figure}[t!] 
\begin{center}
 \includegraphics[width=.99\columnwidth]{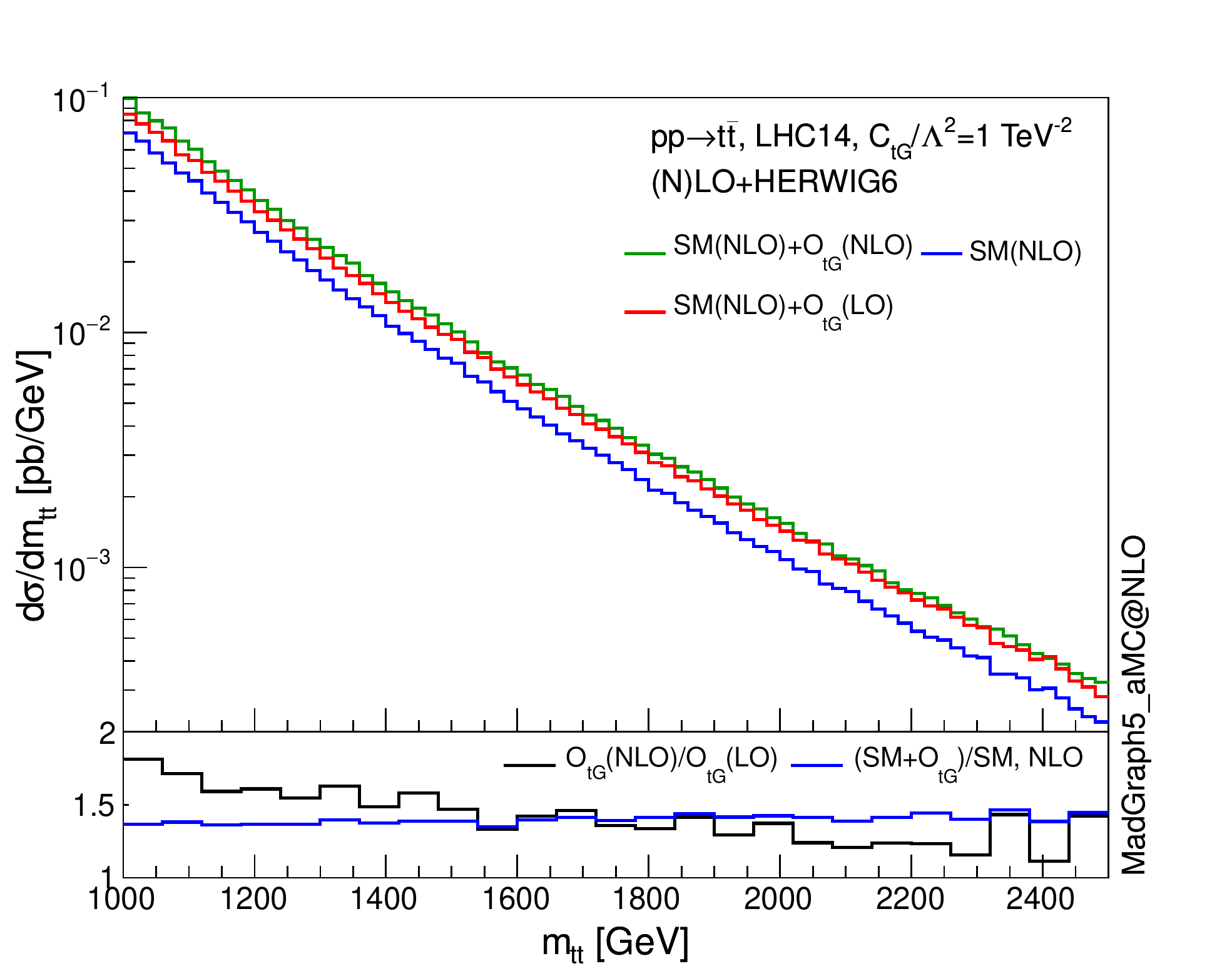}
 \caption{Top quark pair invariant mass distribution above 1 TeV, at the LHC 14
   TeV energy.  The renormalization and factorization scales are taken to be 1
   TeV.  The lower panel shows that the K factor for $O_{tG}$ decreases at high
 mass region, and the signal over background ratio, $\sigma(O_{tG})/\sigma(SM)$,
 is almost a constant.}
\label{fig:mtt}
\end{center}
\end{figure}

In \fig{fig:mtt} we show the top quark pair invariant mass distribution for LHC
at 14 TeV, at high mass region above 1 TeV.  In this calculation we have set
$\mu=1$ TeV. It has been suggested that
kinematical cuts on the $t\bar{t}$ invariant mass can improve the sensitivity
to access the $O_{tG}$ operator, both in the angular
distributions \cite{Bernreuther:2013aga} and in the total cross sections
\cite{Aguilar-Saavedra:2014iga}.  This is because higher momentum transfer is
favoured by the dipole structure of $O_{tG}$.  However, using NLO (or NNLO)
prediction for the SM together with only LO prediction for $O_{tG}$ may lead to
overestimate this effect, because the K factor decreases at larger energy
scales.  From \fig{fig:mtt} we can see that the signal excess,
$\sigma(O_{tG})/\sigma({\rm SM})$, is flat at a large energy range.  In fact, at
order $\mathcal{O}(\Lambda^{-2})$, $\sigma(O_{tG})$ is suppressed by a constant
factor of $m_t^2/\Lambda^2$ compared with $\sigma({\rm SM})$, instead of
$s/\Lambda^2$ or $\sqrt{s}m_t/\Lambda^2$ as one might have expected naively by
power counting.  This is because the Higgs field in $O_{tG}$ always takes the
vacuum expectation value, and the $O_{tG}$ operator flips the chirality of the
top quark in its interference with the SM amplitude. Including higher order
terms in $1/\Lambda^{2}$ can give rise to additional contributions that will
indeed rise faster at large $s$, but if such an effect is large, it would
imply the breakdown of the effective operator framework since the expansion
in $1/\Lambda$ does not converge at large energy.  In \tab{tab:mtt} we show
K factors for SM and $O_{tG}$ as well as $\sigma(O_{tG})/\sigma(SM)$, with no
cuts and with cuts $m_{t\bar t}>1$ TeV and 2 TeV.Both LO and NLO give similar
results for the signal excess, but using LO prediction for $O_{tG}$ with NLO
for SM leads to an artificial rise at large $m_{t\bar t}$, due to the
decreasing K factor of the SM.

\begin{table}
\begin{tabular}{|l|c|c|c|}
\hline
&no cuts & $m_{t\bar t}>1$ TeV & $m_{t\bar t}>2$ TeV
\\\hline
%SM LO [pb] & 598 & 21.5 & 0.705
%\\\hline
%SM NLO [pb] & 893 & 24.9 & 0.544
%\\\hline
%$O_{tG}$ LO [pb] & 192 & 6.06 & 0.204
%\\\hline
%$O_{tG}$ NLO [pb] & 286 & 6.92 & 0.142
%\\\hline
K (SM) & 1.49 & 1.16 & 0.77
\\\hline
K ($O_{tG}$) & 1.49 & 1.14 & 0.69
\\\hline
 $O_{tG}$(LO)/SM(LO) & 0.32 & 0.28 & 0.29
\\\hline
 $O_{tG}$(NLO)/SM(NLO) & 0.32 & 0.28 & 0.26
\\\hline
 $O_{tG}$(LO)/SM(NLO) & 0.21 & 0.24 & 0.37
\\\hline
\end{tabular}
\caption{K factor and signal excess $\sigma(O_{tG})/\sigma(SM)$, with no cuts, 
and with cuts $m_{t\bar t}>1$ TeV and 2 TeV. $\mu=m_t$.  Both LO and
NLO give similar results for the signal excess, but using LO prediction for
$O_{tG}$ and NLO for SM leads to an artificial rise at large $m_{t\bar t}$.
\label{tab:mtt}}
\end{table}

Another interesting observable is the forward-backward asymmetry, $A_{FB}$, 
which has been observed at the Tevatron both by D0 \cite{Abazov:2014cca}
and CDF \cite{Aaltonen:2012it}.  $A_{FB}$ is defined as the asymmetry with respect
to $\Delta y=y_t-y_{\bar t}$.  In the SM the first non-zero contribution arises
at NLO in QCD.  In a dimension-six Lagrangian, only four-fermion operators can
give a contribution at the tree level, and so $A_{FB}$ is another important
observable that distinguishes between different new physics scenarios.  In this
respect, it is useful to know the first non-vanishing contribution from the
CMDM operator, which appears at NLO, to at least have some estimation of the
corresponding theoretical uncertainty related to this quantity.  In our
framework this calculation is straightforward.  We
expand the numerator and the denominator of $A_{FB}$ to NNLO for the SM part
and NLO for the $O_{tG}$ part, and we find
\begin{flalign}
	A_{FB}=&\frac{N_{\rm EW}+\alpha_s^3N_3+\alpha_s^4N_4
	+\alpha_s^3\frac{C_{tG}}{\Lambda^2} N_{tG}
	+\mathcal{O}(\alpha_s^5,\alpha_s^4\Lambda^{-2}) }
	{\alpha_s^2D_2+\alpha_s^3D_3+\alpha_s^4D_4
	+\alpha_s^3\frac{C_{tG}}{\Lambda^2} D_{tG}
		+\mathcal{O}(\alpha_s^5,\alpha_s^4\Lambda^{-2})}
		\nonumber\\
	=&A_{FB}(SM)+\frac{C_{tG}}{\Lambda^2}\frac{\alpha_s N_{tG}}{D_2}
	+\mathcal{O}(\alpha_s^3,\alpha_s^2\Lambda^{-2})
	\nonumber\\
	=&0.095\pm0.007+C_{tG}\,0.021^{+0.003}_{-0.002}\left( \frac{\rm TeV}{\Lambda} \right)^2 
\end{flalign}
where the SM prediction at NNLO in QCD is taken from
Ref.~\cite{Czakon:2014xsa}, and the uncertainties of the second term come from
scale variation.  We thus expect a small modification to $A_{FB}$ from a
non-vanishing CMDM.  Given the current limit on $C_{tG}$, however, this
contribution is much smaller than the experiment uncertainties.

Once differential cross sections are known, one can consider constraining the
CMDM by using the normalized distributions of the $t\bar t$ observables.
Unfortunately from \fig{fig:mtt8} and \fig{fig:pt_t} one can see that the
shapes of the distributions from $O_{tG}$ are not significantly different than
those from the SM.  As a result the limits obtained only by using the shape of the
distribution will be loose.  As an example we consider the $t\bar t$ invariant
mass distribution at 7 TeV measured by the CMS collaboration \cite{Chatrchyan:2012saa}.
We take only the first four bins, i.e.~from 345 to 650 GeV, to ensure the
validity of the expansion in $1/\Lambda^2$ for $\Lambda$ around TeV scale.  We
perform a simple $\chi^2$ fit
for the differential cross section normalized within these four bins.  We add
the experimental and theoretical errors in quadrature.  The SM prediction is
computed with {\sc MadGraph5\_aMC@NLO} and then normalized to the most accurate
NNLO+NNLL prediction \cite{Czakon:2013goa}, with uncertainties coming from the
renormalization and factorization scale variation.  The 95\% CL allowed region
is [-5.0,12.8], using LO predictions for the $O_{tG}$ contribution,
and [-0.6,10.9] when using NLO prediction.  Despite a significant improvement
at NLO, the limit itself is much looser than those obtained from total cross
sections, therefore we expect only a small improvement when the distribution
information is combined together with the total cross section in such analyses.

\subsection{Decayed top quarks}

The above results indicate that the kinematic observables constructed from
stable $t\bar t$ system are not very sensitive to the size of the top-quark CMDM.
We therefore move on to include top-quark decays, where we expect that the decay
products preserve the spin information of the top quarks, and thus can be more
sensitive to the dipole structure in the operator. As an example we focus on
the dimuon channel, where both top quarks decay semileptonically into a $b$-quark
and a muon. We use the
{\sc MadSpin} package \cite{Artoisenet:2012st} to decay the top quarks, so that
the spin correlation at LO accuracy is preserved in the simulation.

\begin{figure}[t] 
\begin{center}
 \includegraphics[width=.99\columnwidth]{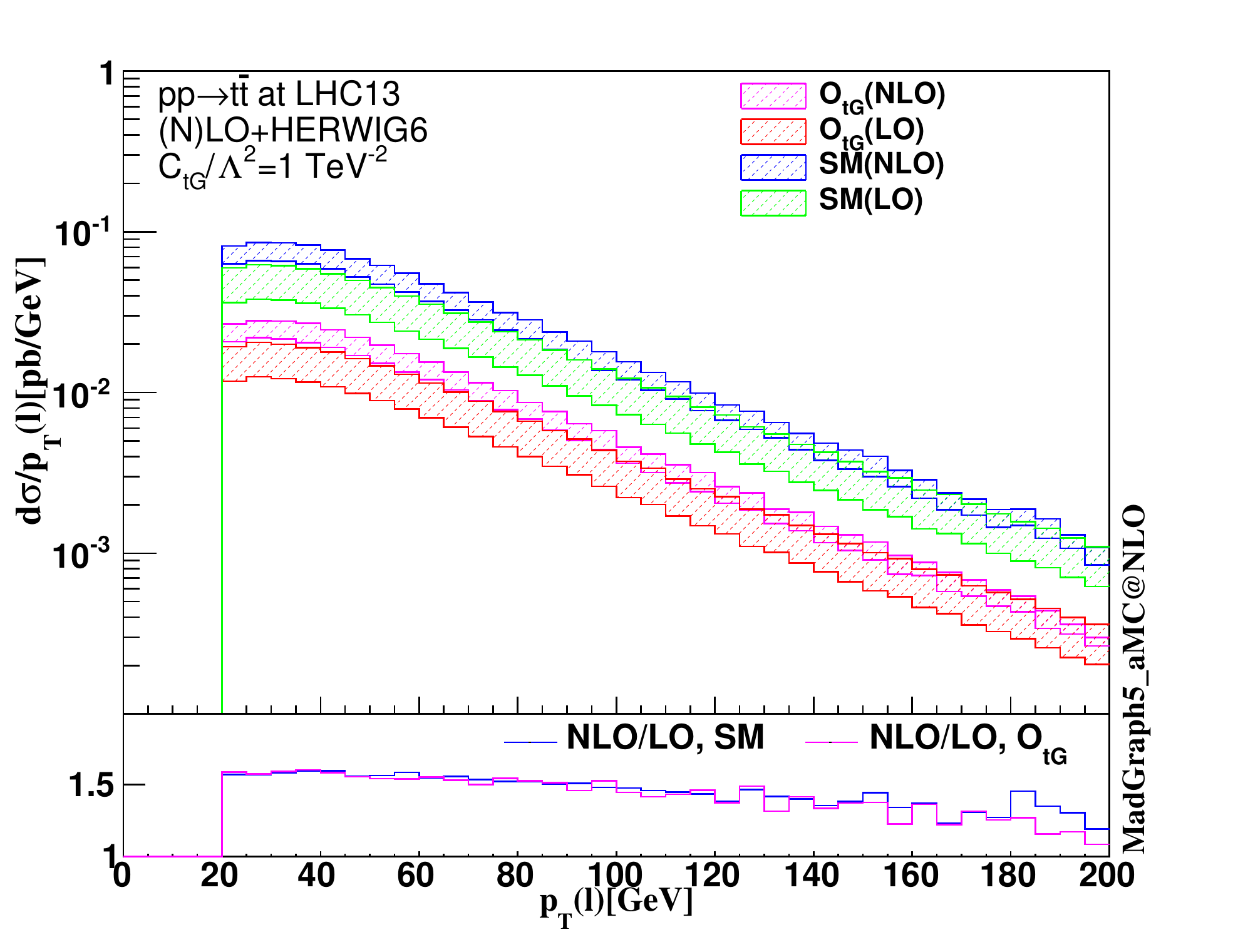}
 \caption{Transverse momentum distribution of the hardest muon at LHC 13 TeV.}
\label{fig:lpt}
\end{center}
\end{figure}

\begin{figure}[t] 
\begin{center}
 \includegraphics[width=.99\columnwidth]{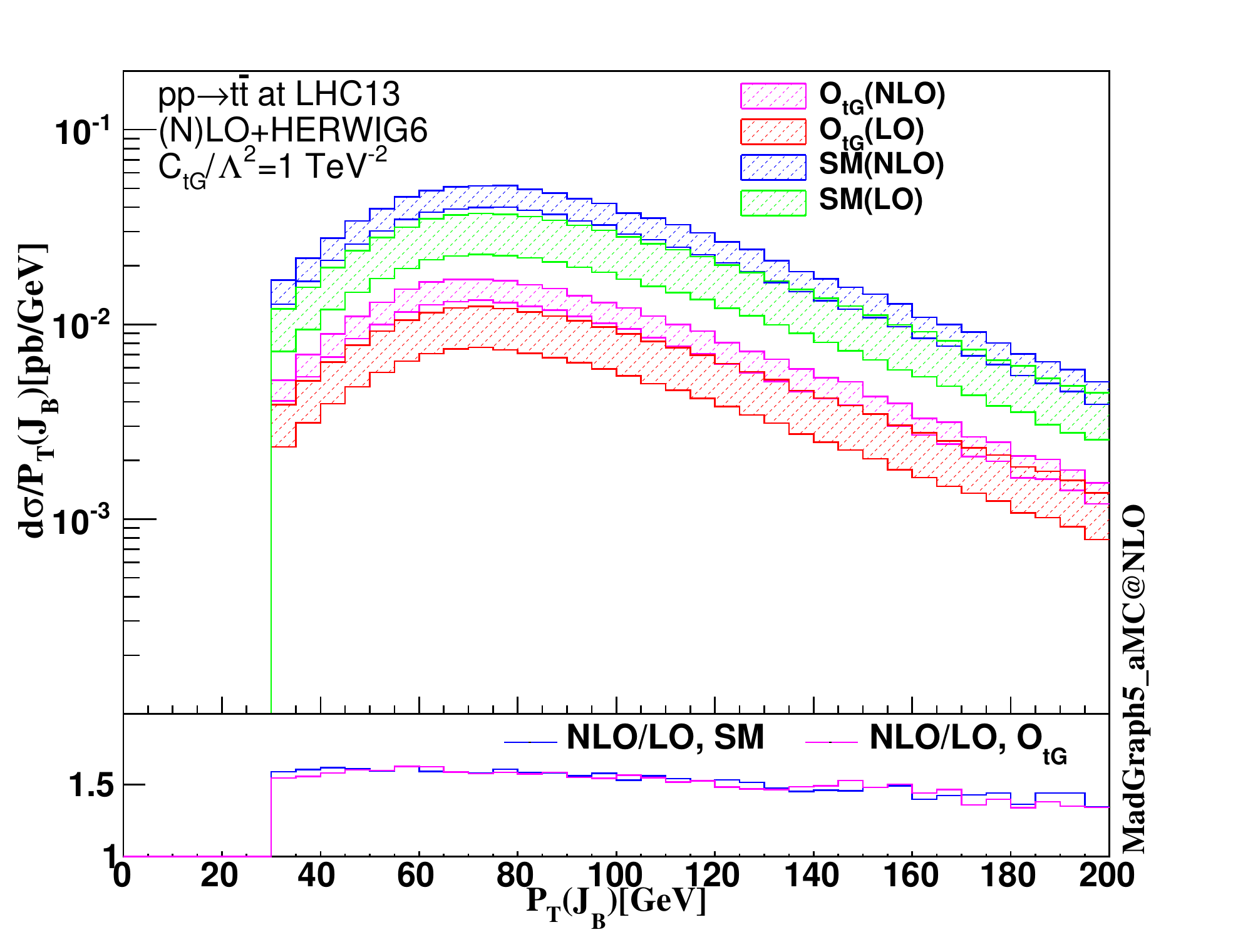}
 \caption{Transverse momentum distribution of the hardest $b$-jet at LHC 13 TeV.}
\label{fig:bj1pt}
\end{center}
\end{figure}

\begin{figure}[t] 
\begin{center}
 \includegraphics[width=.99\columnwidth]{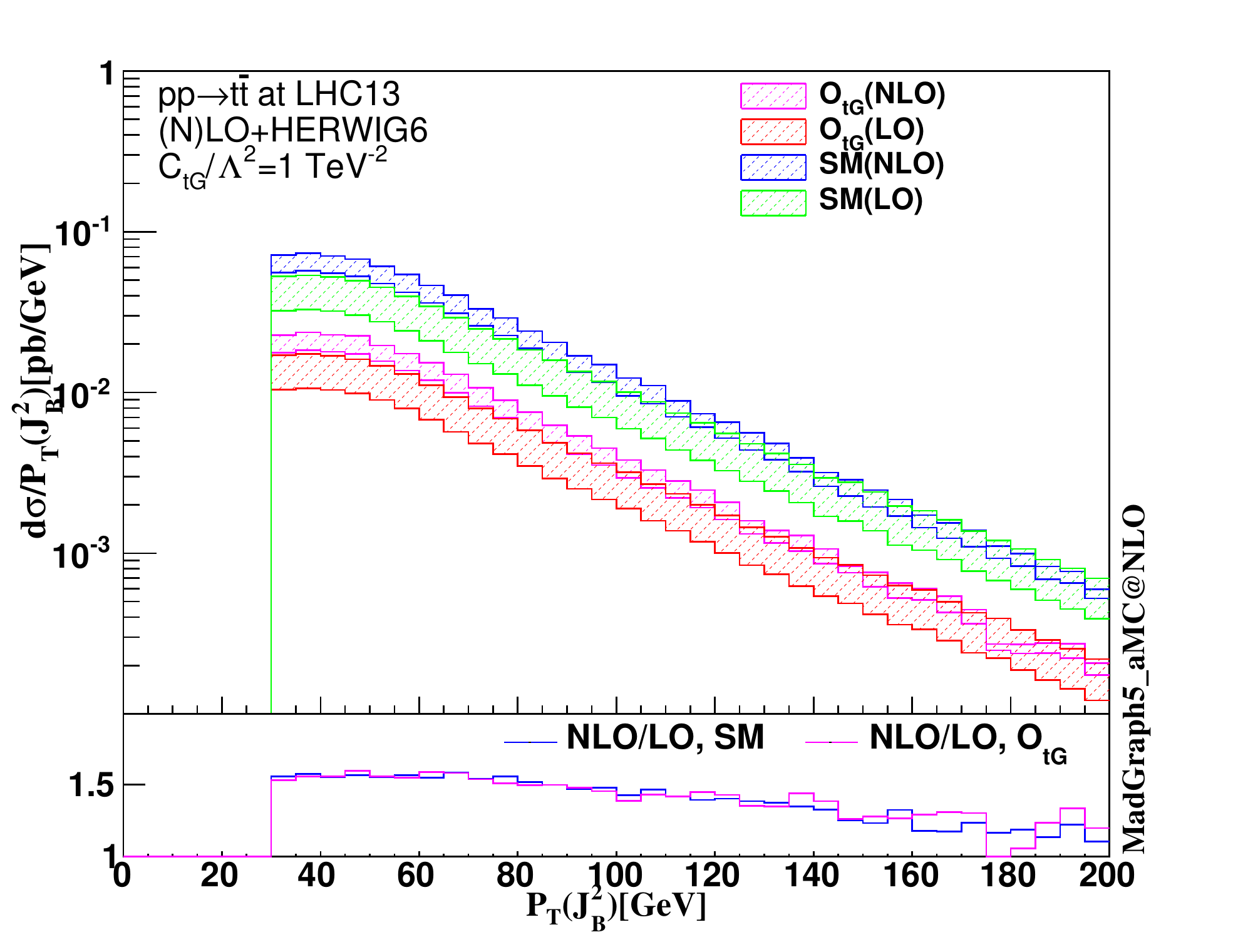}
 \caption{Transverse momentum distribution of the second hardest $b$-jet at LHC 13 TeV.}
\label{fig:bj2pt}
\end{center}
\end{figure}

\begin{figure}[t] 
\begin{center}
 \includegraphics[width=.99\columnwidth]{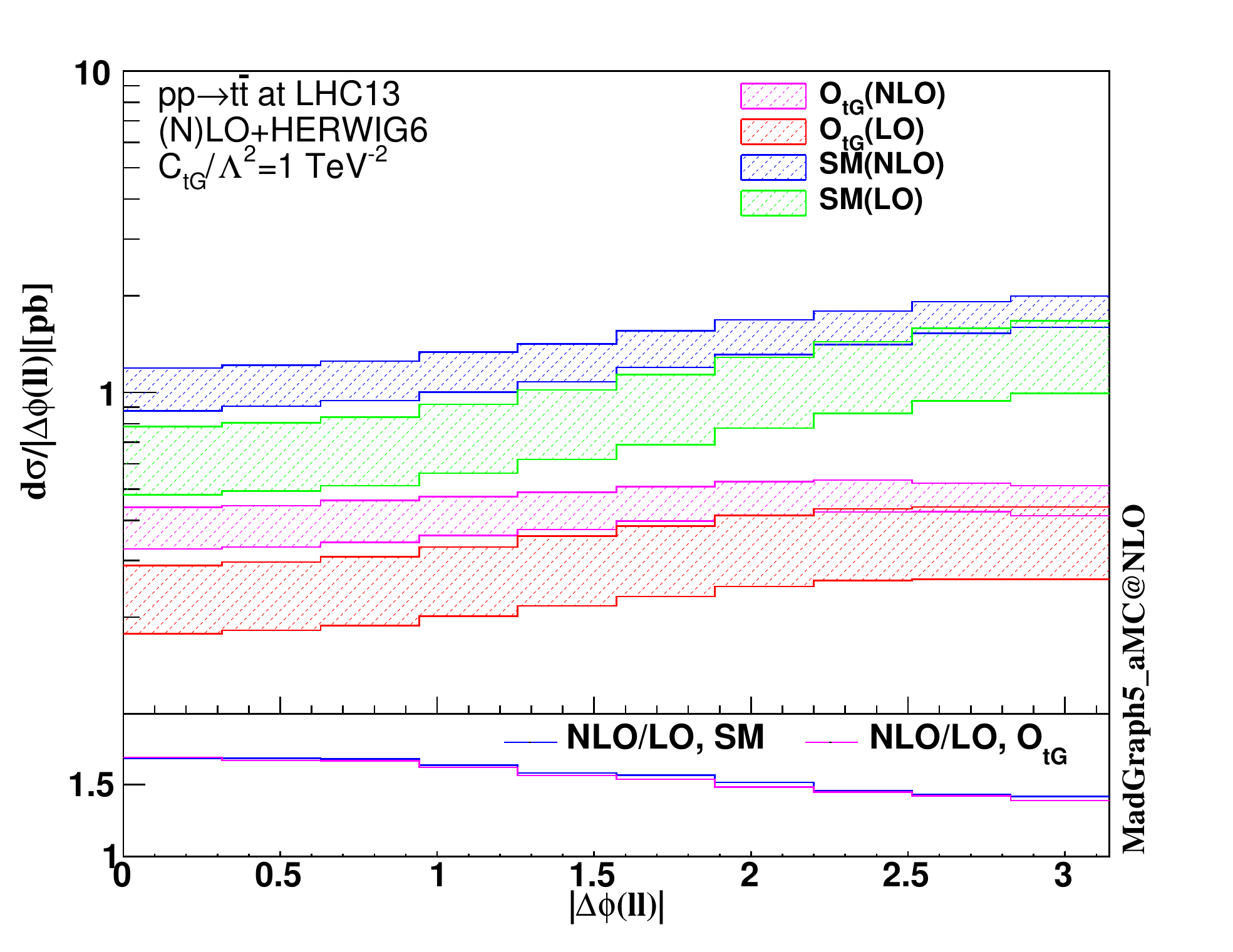}
 \caption{Difference in azimuthal angle between the two selected muons at LHC 13 TeV.
	 Note that the K factor changes in a way that enhances the deviation of 
	 the $O_{tG}$ contribution from the SM.}
\label{fig:llazi}
\end{center}
\end{figure}

In our simulation we use the anti-$k_T$ algorithm for the jets with radius
$R=0.5$.  The following cuts are imposed to mimic the environment of a real
detector: $p_T(j)>30\GeV$, $|\eta(j)|<2.5$, $p_T(\ell)>20\GeV$ and
$\eta(\ell)|<2.5$, where $j$ refers to jets and $\ell$ to muons. 
At least two jets, from which at least one containing a
$b$-hadron, and exactly one pair of isolated muons are required. The
isolation criteria is achieved by imposing a maximum value of 0.15 on the ratio
of the scalar sum of $p_T$ of all hadronic tracks within $\Delta
R=\sqrt{\Delta\eta^2+\Delta\phi^2}<0.3$ around the muon candidate, to the
transverse momentum of the muon.  We show in \fig{fig:lpt} the transverse
momentum distribution of the muon. In \figs{fig:bj1pt}{fig:bj2pt} we show 
the hardest (largest $p_T$) and the second hardest
(when present in the event)  $b$-jets, respectively.
In \fig{fig:llazi} the azimuthal angle difference between
the two selected muons is shown. 

This last distribution is particularly important for the measurement of
the $O_{tG}$ operator, because it is sensitive to the spin correlation
between the top quarks.  The anomalous top-quark CMDM affects the spin
correlation of the $t\bar t$ system \cite{Bernreuther:2013aga}, and its effects
have been searched for by the CMS collaboration \cite{CMS:2014bea}. 
The contribution from the $O_{tG}$ operator
to this distribution, expanded linearly in $\CtG$, takes the following form
\be
\left(\frac{1}{\sigma}\frac{d\sigma}{d|\Delta\phi|}\right)=
\left(\frac{1}{\sigma}\frac{d\sigma}{d|\Delta\phi|}\right)_{\rm SM}
+\frac{C_{tG}}{\Lambda^2}\left(\frac{1}{\sigma}\frac{d\sigma}{d|\Delta\phi|}\right)_{\rm
NP}
\label{eq:deltaphi}
\ee
which is valid provided $C_{tG}\beta_1/\Lambda^2\ll \sigma_{SM}$. 
In the spirit of perturbation theory, the second term
on the r.h.s.~of \eq{eq:deltaphi} can be expanded to ${\cal O}(\alpha_S)$:
\begin{equation}
	\left(\frac{1}{\sigma}\frac{d\sigma}{d|\Delta\phi|}\right)^{\rm NLO}_{\rm NP}=
	\left(\frac{1}{\sigma}\frac{d\sigma}{d|\Delta\phi|}\right)^{\rm LO}_{\rm NP}+
	\left(\frac{1}{\sigma}\frac{d\sigma}{d|\Delta\phi|}\right)^{(1)}_{\rm NP}
\label{eq:deltaphiNP}
\end{equation}
where
\be
\left(\frac{1}{\sigma}\frac{d\sigma}{d|\Delta\phi|}\right)^{\rm LO}_{\rm NP}=
\frac{1}{\sigma^{\rm LO}_{\rm SM}}\left( \frac{d\sigma^{\rm
LO}_{O_{tG}}}{d|\Delta\phi|}  \right)-
\frac{\beta_1^{\rm LO}}{\sigma^{\rm LO\,2}_{\rm SM}}\left(\frac{d\sigma^{\rm
LO}_{\rm SM}}{d|\Delta\phi|}  \right)
\label{eq:deltaphiNPLO}
\ee
and
\begin{widetext}
\begin{flalign}
\left(\frac{1}{\sigma}\frac{d\sigma}{d|\Delta\phi|}\right)^{(1)}_{\rm NP}&=
\frac{1}{\sigma^{\rm LO}_{\rm SM}}\left( \frac{d\sigma^{(1)}_{O_{tG}}}{d|\Delta\phi|}  \right)-
\frac{1}{\sigma^{\rm LO\,2}_{\rm SM}}\left[\beta_1^{(1)}\left(\frac{d\sigma^{\rm LO}_{\rm SM}}{d|\Delta\phi|}\right)+
\beta_1^{\rm LO}\left(\frac{d\sigma^{(1)}_{\rm SM}}{d|\Delta\phi|}\right)+
\sigma^{(1)}_{\rm SM}\left(\frac{d\sigma^{\rm LO}_{O_{tG}}}{d|\Delta\phi|}\right)\right]
%\nonumber\\ &
+ \frac{2\sigma^{(1)}_{\rm SM}\beta_1^{\rm LO}}{\sigma^{\rm LO\,3}_{\rm SM}}\left(\frac{d\sigma^{\rm LO}_{\rm SM}}{d|\Delta\phi|}\right)
\label{eq:deltaphiNPNLO}
\end{flalign}
\end{widetext}
where $d\sigma_{O_{tG}}$ represents the distribution from operator $O_{tG}$
for $C_{tG}/\Lambda^2=1$ TeV$^{-2}$. The superscript $(1)$ indicates the 
$\alpha_S$ correction to the corresponding LO quantity. 

\begin{figure}[t] 
\begin{center}
 \includegraphics[width=.99\columnwidth]{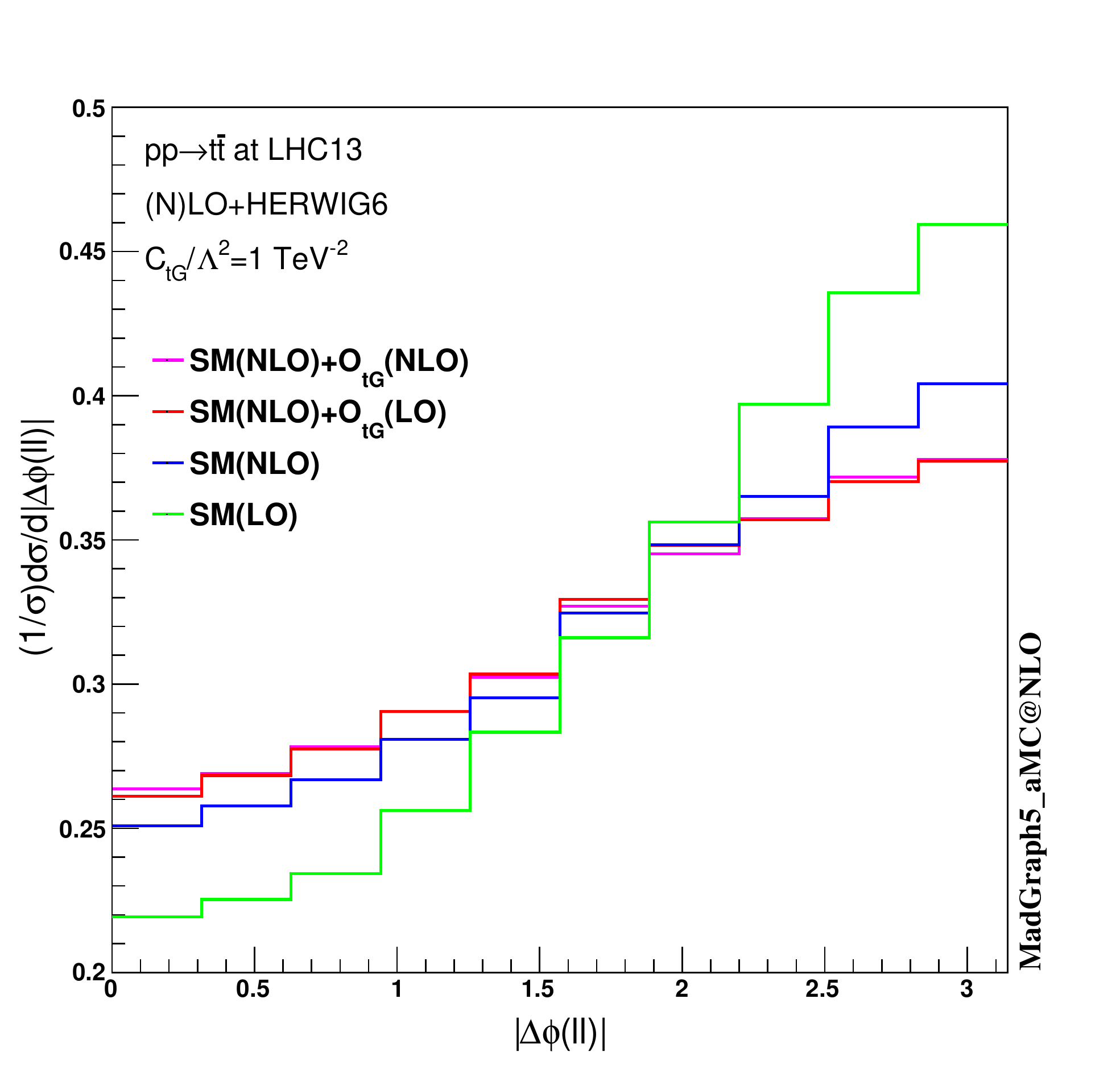}
 \caption{Normalized distributions of the difference in azimuthal angle between muons.}
\label{fig:deltaphi}
\end{center}
\end{figure}

\begin{figure}[t] 
\begin{center}
 \includegraphics[width=.99\columnwidth]{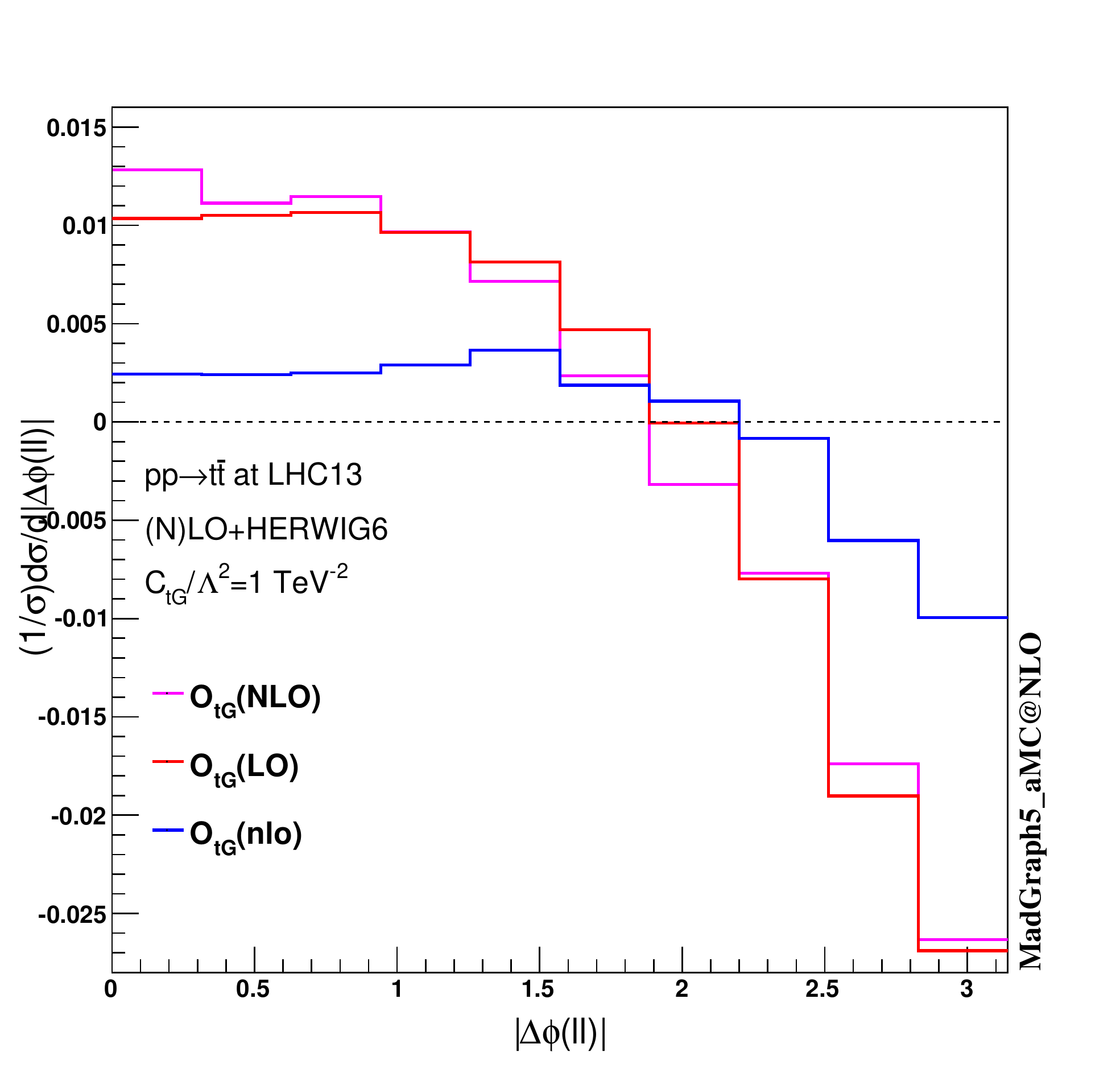}
 \caption{New physics contribution to the normalized dimuon distribution at the LHC (13 TeV), $(\sigma^{-1}d\sigma/d|\Delta\phi|)_{NP}$ defined in \eqs{eq:deltaphi}{eq:deltaphiNPnlo}.}
\label{fig:deltaphiNP}
\end{center}
\end{figure}

The predicted distribution is shown in \fig{fig:deltaphi} for
$C_{tG}/\Lambda^2=1$ TeV$^{-2}$. The $O_{tG}$ contribution has a peculiar
structure which tends to flatten the distribution. In \fig{fig:deltaphiNP} we
show the $O_{tG}$ distributions solely, as defined in
\eqs{eq:deltaphiNP}{eq:deltaphiNPNLO}. 
The first observation is that the purple and the red curves are very close to
each other, indicating that the LO and NLO results are very similar.  The reason
is that we are plotting the normalized distribution, and since the K factors
are almost the same for the SM and for the $O_{tG}$, they cancel each other
when taking the ratio.  In fact, one can see that \eq{eq:deltaphiNPLO} vanishes
if the K factor is a constant.  Alternatively, if we use NLO prediction for
the SM but only LO prediction for $O_{tG}$, following the logic that the
radiative correction on the new physics effect is of higher order, then we will
have
\be
\left(\frac{1}{\sigma}\frac{d\sigma}{d|\Delta\phi|}\right)^{\rm nlo}_{\rm NP}=
\frac{1}{\sigma^{\rm NLO}_{\rm SM}}\left( \frac{d\sigma^{\rm
LO}_{O_{tG}}}{d|\Delta\phi|}  \right)-
\frac{\beta_1^{\rm LO}}{\sigma^{\rm NLO\,2}_{\rm SM}}\left(\frac{d\sigma^{\rm
LO}_{\rm SM}}{d|\Delta\phi|}  \right)
\label{eq:deltaphiNPnlo}
\ee
This result, after expanding in $\alpha_s$, contains only part of the
$\mathcal{O}(\alpha_s\Lambda^{-2})$ corrections in \eq{eq:deltaphiNPNLO} (and
so we refer to as ``nlo''). They
come from the $\mathcal{O}(\alpha_s)$ corrections to the normalization, but not
directly to the $O_{tG}$ contribution.  The missing
$\mathcal{O}(\alpha_s\Lambda^{-2})$ terms actually make a large difference, as
illustrated by the blue curve in \fig{fig:deltaphiNP}.  One can see that
\eq{eq:deltaphiNPnlo} gives a much lower estimation for the effect of $O_{tG}$.
This is not only because of the overall size of the K factor, but also due to
the fact that the K factor is a decreasing function of $\Delta\phi(ll)$, and so
the way it changes adds coherently to the difference in shapes between $O_{tG}$
and SM distributions, as can be seen in \fig{fig:llazi}.  As a result, using
NLO prediction for the SM together with only LO
prediction for $O_{tG}$ significantly underestimates the power of
$\Delta\phi(ll)$ in discriminating the $O_{tG}$ contribution from the SM.  Also
note that, the fact that \eq{eq:deltaphiNPnlo} and the LO prediction in
\eq{eq:deltaphiNPLO} differ implies that the LO prediction for $O_{tG}$
has a large uncertainty due to the missing $\mathcal{O}(\alpha_s\Lambda^{-2})$
terms, which turn out to have a large effect in this special case.
Thus our work improves the precision level of this prediction by completing
the missing $\mathcal{O}(\alpha_s\Lambda^{-2})$ terms.

\begin{figure}[t] 
\begin{center}
 \includegraphics[width=.99\columnwidth]{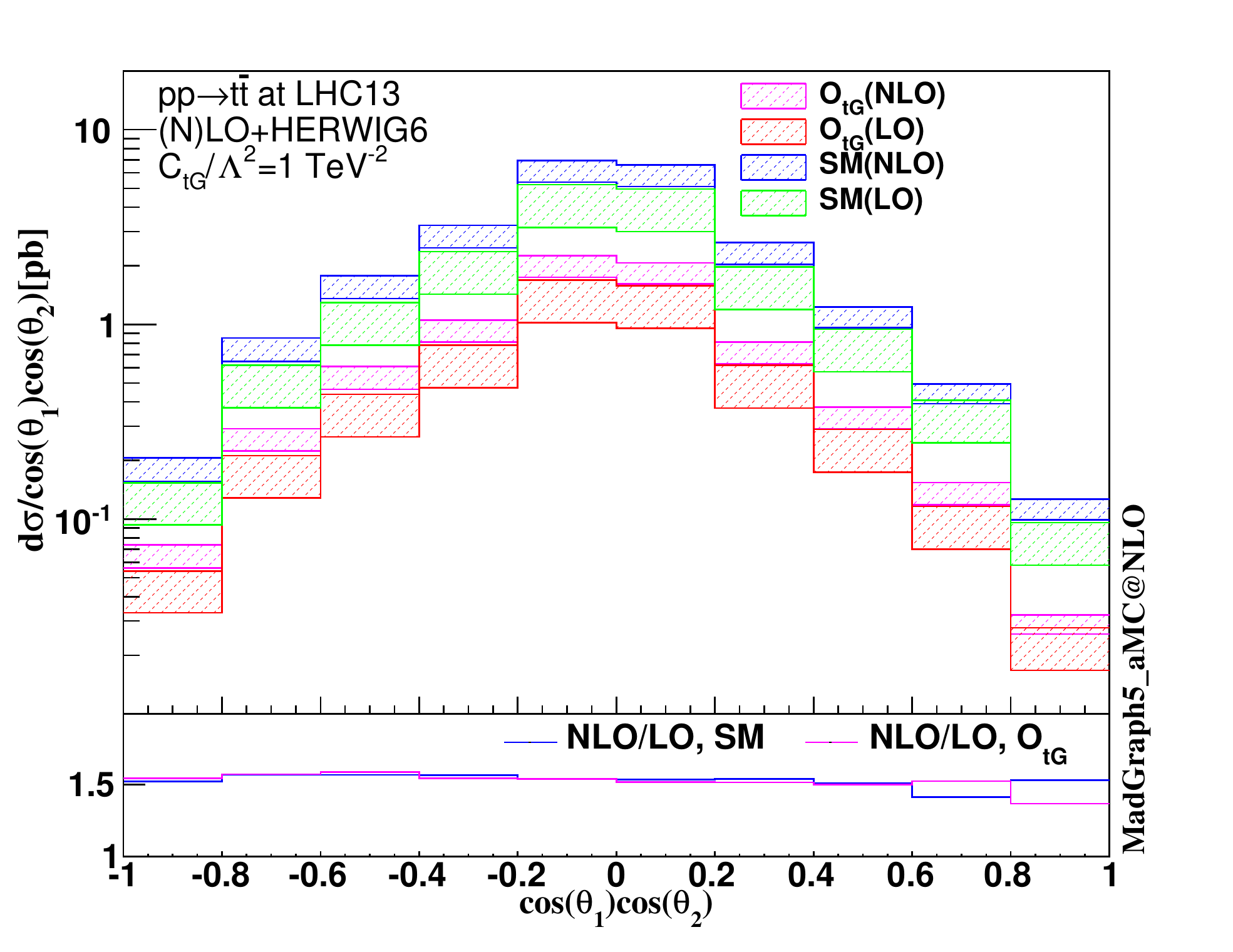}
 \caption{$\cos\theta_1\cos\theta_2$ distribution at LHC 13 TeV.}
\label{fig:cos1cos2}
\end{center}
\end{figure}

\begin{figure}[t!] 
\begin{center}
 \includegraphics[width=.99\columnwidth]{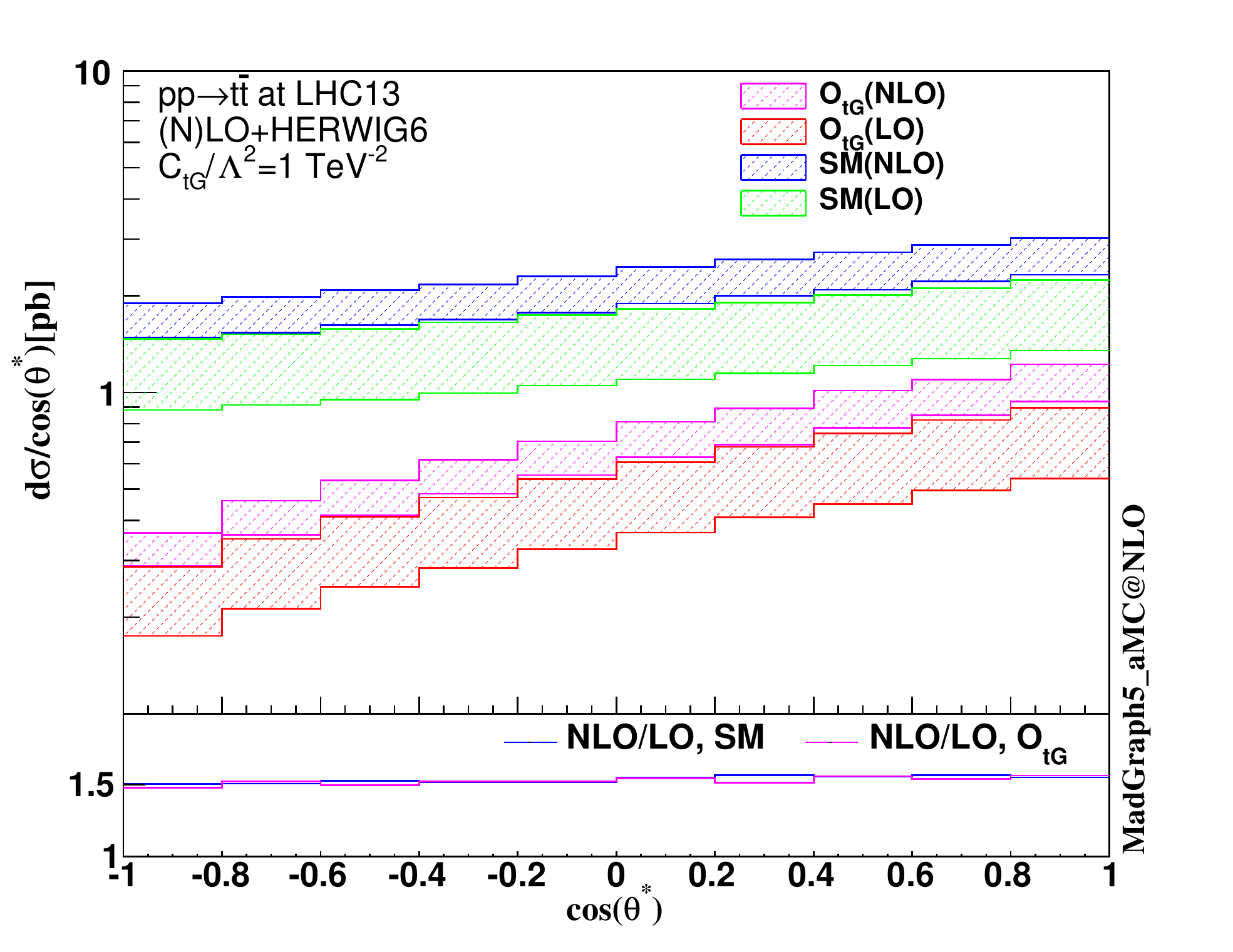}
 \caption{$\cos \theta^*$ distribution at LHC 13 TeV.}
\label{fig:cosll}
\end{center}
\end{figure}

We also show for completeness two more angular distributions which have been
studied by Ref.~\cite{Bernreuther:2013aga}. Following
Ref.~\cite{Bernreuther:2013aga} we define $\vec{\ell^-}$ ($\vec{\ell^+}$) as
the
momenta of the (anti-)muon in the rest frame of anti-top (top) quarks and
$\vec{k}$ ($\bar{\vec{k}}$) the momenta of the top (anti-top) in the
zero-mometum frame. In \fig{fig:cos1cos2} we show the distribution of
$\cos\theta_1\cos\theta_2$, where $\theta_1$ ($\theta_2$) is the angle
$\measuredangle( \vec{\ell^-},\bar{\vec{k}})$ ($\measuredangle
(\vec{\ell^+},\vec{k})$).  In \fig{fig:cosll} we show the normalised
distribution of $\cos \theta^*$, where $\theta^*$ is the angle $\measuredangle
(\vec{\ell^-},\vec{\ell^+})$.  Contrary to the $|\Delta\phi(\ell\ell)|$ case,
where the QCD corrections enhances the anomalous coupling contribution
to the shape, in these cases, we observe a uniform QCD correction with no
effect in the normalised distributions.

\section{Conclusions}
\label{sec:6}

In this work we have presented the NLO calculation for top-quark pair production,
including an anomalous top-quark CMDM, as described by the dimension-six operator
$O_{tG}$.  Our calculation is implemented in the {\sc MadGraph5\_aMC@NLO}
framework, which allows the result to be matched to parton shower automatically.
We have studied the impact of QCD corrections to the contribution of the CMDM
in top quark pair production, both on total cross section as well as on various
distributions.   

The QCD correction increases the overall contribution from the $O_{tG}$ operator. 
For the total cross section for example, the increase is, at central scale,
12\%, 43\% and 48\% for Tevatron, LHC8 and LHC14 respectively. Moreover, 
the NLO calculation significantly reduces the scale uncertainty of the
contribution from $O_{tG}$. Limits on the coefficient of $C_{tG}$ are therefore
improved.  Our predicted allowed range at 95\% CL using Tevatron and LHC8 data
is $-0.32<\CtG<0.30$ (assuming $\Lambda=1\TeV$), which in terms of $d_V$
parameter gives $-0.0096 < d_V < 0.0090$. 

Our implementation can be used for various exclusive studies.  We have shown
representative distributions for both stable and decayed top quarks as
examples.  We observed a significant reduction of scale variation in all
distributions.  The differential K factor is not a constant, but for all
observables we have studied, it is similar to the SM K factor.  Therefore we
expect that using the SM K factor to rescale the LO contribution of $O_{tG}$
can be a good approximation for an NLO prediction in most cases.  On the other
hand, using NLO SM prediction together with LO prediction of $O_{tG}$ can be
misleading in analysis where the ratio between $O_{tG}$ contribution and SM
contribution can play a role.  Observables sensitive to spin correlation can
also be studied in the same framework, provided that the {\sc MadSpin} package
is used to preserve the spin information of the top quarks.  This is
particularly useful for spin correlation measurements where 
limits can be set by using various angular distributions of
the decay products.
We showed that the NLO correction
does not significantly change the LO prediction, but instead it increases
the precision level, in particular for the $\Delta\phi(ll)$ distribution,
where using NLO SM prediction together with LO prediction of $O_{tG}$
can lead to underestimate the effect from top-quark CMDM.

Our theoretical approach is based on the effective field theory for top-quark
couplings, and is a first step of the automation of the top-quark flavor-diagonal
operators in the {\sc MadGraph5\_aMC@NLO} framework.  The next step is to extend
our study to other top-quark operators, including the CP-odd ones such as the CEDM, 
as well as other electroweak couplings of the top quark.
These studies will pave the way to a global analysis for top quark couplings
using the effective field theory framework.

\section{Acknowledgements}
We would like to thank C.~Degrande, V.~Hirschi, F.~Maltoni and M.~Zaro for many
helpful discussions and patient explanations.
C.~Z.~is supported by U.S.~Department of Energy under Grant
No.~DE-AC02-98CH10886.
D.~B.~F.~is supported by the Danish National Research Foundation, grant number DNRF90. 

\bibliography{OtG}
\bibliographystyle{apsrev4-1_title}

\end{document}